\documentstyle[multicol,aps,epsf,rotate]{revtex}

\begin{document}
\def\PsfigVersion{1.9}
\ifx\undefined\psfig\else \fi

%

\let\LaTeXAtSign=\@
\let\@=\relax
\edef\psfigRestoreAt{\catcode`\@=\number\catcode`@\relax}
\catcode`\@=11\relax
\newwrite\@unused
\def\ps@typeout#1{{\let\protect\string\immediate\write\@unused{#1}}}
\ps@typeout{psfig/tex \PsfigVersion}


\def\figurepath{./}
\def\psfigurepath#1{\edef\figurepath{#1}}

%
%
\def\@nnil{\@nil}
\def\@empty{}
\def\@psdonoop#1\@@#2#3{}
\def\@psdo#1:=#2\do#3{\edef\@psdotmp{#2}\ifx\@psdotmp\@empty \else
    \expandafter\@psdoloop#2,\@nil,\@nil\@@#1{#3}\fi}
\def\@psdoloop#1,#2,#3\@@#4#5{\def#4{#1}\ifx #4\@nnil \else
       #5\def#4{#2}\ifx #4\@nnil \else#5\@ipsdoloop #3\@@#4{#5}\fi\fi}
\def\@ipsdoloop#1,#2\@@#3#4{\def#3{#1}\ifx #3\@nnil 
       \let\@nextwhile=\@psdonoop \else
      #4\relax\let\@nextwhile=\@ipsdoloop\fi\@nextwhile#2\@@#3{#4}}
\def\@tpsdo#1:=#2\do#3{\xdef\@psdotmp{#2}\ifx\@psdotmp\@empty \else
    \@tpsdoloop#2\@nil\@nil\@@#1{#3}\fi}
\def\@tpsdoloop#1#2\@@#3#4{\def#3{#1}\ifx #3\@nnil 
       \let\@nextwhile=\@psdonoop \else
      #4\relax\let\@nextwhile=\@tpsdoloop\fi\@nextwhile#2\@@#3{#4}}
%
\ifx\undefined\fbox
\newdimen\fboxrule
\newdimen\fboxsep
\newdimen\ps@tempdima
\newbox\ps@tempboxa
\fboxsep = 3pt
\fboxrule = .4pt
\long\def\fbox#1{\leavevmode\setbox\ps@tempboxa\hbox{#1}\ps@tempdima\fboxrule
    \advance\ps@tempdima \fboxsep \advance\ps@tempdima \dp\ps@tempboxa
   \hbox{\lower \ps@tempdima\hbox
  {\vbox{\hrule height \fboxrule
          \hbox{\vrule width \fboxrule \hskip\fboxsep
          \vbox{\vskip\fboxsep \box\ps@tempboxa\vskip\fboxsep}\hskip 
                 \fboxsep\vrule width \fboxrule}
                 \hrule height \fboxrule}}}}
\fi
%
%
\newread\ps@stream
\newif\ifnot@eof       
\newif\if@noisy        
\newif\if@atend        
\newif\if@psfile       
%
%
{\catcode`\%=12\global\gdef\epsf@start{
\def\epsf@PS{PS}
\def\epsf@getbb#1{%
%
%
\openin\ps@stream=#1
\ifeof\ps@stream\ps@typeout{Error, File #1 not found}\else
%
%
   {\not@eoftrue \chardef\other=12
    \def\do##1{\catcode`##1=\other}\dospecials \catcode`\ =10
    \loop
       \if@psfile
	  \read\ps@stream to \epsf@fileline
       \else{
	  \obeyspaces
          \read\ps@stream to \epsf@tmp\global\let\epsf@fileline\epsf@tmp}
       \fi
       \ifeof\ps@stream\not@eoffalse\else
%
%
       \if@psfile\else
       \expandafter\epsf@test\epsf@fileline:. \\%
       \fi
%
%
          \expandafter\epsf@aux\epsf@fileline:. \\%
       \fi
   \ifnot@eof\repeat
   }\closein\ps@stream\fi}%
%
%
\long\def\epsf@test#1#2#3:#4\\{\def\epsf@testit{#1#2}
			\ifx\epsf@testit\epsf@start\else
\ps@typeout{Warning! File does not start with `\epsf@start'.  It may not be a PostScript file.}
			\fi
			\@psfiletrue} 
%
%
{\catcode`\%=12\global\let\epsf@percent=
%
%
%
\long\def\epsf@aux#1#2:#3\\{\ifx#1\epsf@percent
   \def\epsf@testit{#2}\ifx\epsf@testit\epsf@bblit
	\@atendfalse
        \epsf@atend #3 . \\%
	\if@atend	
	   \if@verbose{
		\ps@typeout{psfig: found `(atend)'; continuing search}
	   }\fi
        \else
        \epsf@grab #3 . . . \\%
        \not@eoffalse
        \global\no@bbfalse
        \fi
   \fi\fi}%
%
%
\def\epsf@grab #1 #2 #3 #4 #5\\{%
   \global\def\epsf@llx{#1}\ifx\epsf@llx\empty
      \epsf@grab #2 #3 #4 #5 .\\\else
   \global\def\epsf@lly{#2}%
   \global\def\epsf@urx{#3}\global\def\epsf@ury{#4}\fi}%
%
%
\def\epsf@atendlit{(atend)} 
\def\epsf@atend #1 #2 #3\\{%
   \def\epsf@tmp{#1}\ifx\epsf@tmp\empty
      \epsf@atend #2 #3 .\\\else
   \ifx\epsf@tmp\epsf@atendlit\@atendtrue\fi\fi}


\chardef\psletter = 11 
\chardef\other = 12

\newif \ifdebug 
\newif\ifc@mpute 
\c@mputetrue 

\let\then = \relax
\def\r@dian{pt }
\let\r@dians = \r@dian
\let\dimensionless@nit = \r@dian
\let\dimensionless@nits = \dimensionless@nit
\def\internal@nit{sp }
\let\internal@nits = \internal@nit
\newif\ifstillc@nverging
\def \Mess@ge #1{\ifdebug \then \message {#1} \fi}

{ 
	\catcode `\@ = \psletter
	\gdef \nodimen {\expandafter \n@dimen \the \dimen}
	\gdef \term #1 #2 #3%
	       {\edef \t@ {\the #1}
		\edef \t@@ {\expandafter \n@dimen \the #2\r@dian}%
		\t@rm {\t@} {\t@@} {#3}%
	       }
	\gdef \t@rm #1 #2 #3%
	       {{%
		\count 0 = 0
		\dimen 0 = 1 \dimensionless@nit
		\dimen 2 = #2\relax
		\Mess@ge {Calculating term #1 of \nodimen 2}%
		\loop
		\ifnum	\count 0 < #1
		\then	\advance \count 0 by 1
			\Mess@ge {Iteration \the \count 0 \space}%
			\Multiply \dimen 0 by {\dimen 2}%
			\Mess@ge {After multiplication, term = \nodimen 0}%
			\Divide \dimen 0 by {\count 0}%
			\Mess@ge {After division, term = \nodimen 0}%
		\repeat
		\Mess@ge {Final value for term #1 of 
				\nodimen 2 \space is \nodimen 0}%
		\xdef \Term {#3 = \nodimen 0 \r@dians}%
		\aftergroup \Term
	       }}
	\catcode `\p = \other
	\catcode `\t = \other
	\gdef \n@dimen #1pt{#1} 
}

\def \Divide #1by #2{\divide #1 by #2} 

\def \Multiply #1by #2
       {{
	\count 0 = #1\relax
	\count 2 = #2\relax
	\count 4 = 65536
	\Mess@ge {Before scaling, count 0 = \the \count 0 \space and
			count 2 = \the \count 2}%
	\ifnum	\count 0 > 32767 
	\then	\divide \count 0 by 4
		\divide \count 4 by 4
	\else	\ifnum	\count 0 < -32767
		\then	\divide \count 0 by 4
			\divide \count 4 by 4
		\else
		\fi
	\fi
	\ifnum	\count 2 > 32767 
	\then	\divide \count 2 by 4
		\divide \count 4 by 4
	\else	\ifnum	\count 2 < -32767
		\then	\divide \count 2 by 4
			\divide \count 4 by 4
		\else
		\fi
	\fi
	\multiply \count 0 by \count 2
	\divide \count 0 by \count 4
	\xdef \product {#1 = \the \count 0 \internal@nits}%
	\aftergroup \product
       }}

\def\r@duce{\ifdim\dimen0 > 90\r@dian \then   
		\multiply\dimen0 by -1
		\advance\dimen0 by 180\r@dian
		\r@duce
	    \else \ifdim\dimen0 < -90\r@dian \then  
		\advance\dimen0 by 360\r@dian
		\r@duce
		\fi
	    \fi}

\def\Sine#1%
       {{%
	\dimen 0 = #1 \r@dian
	\r@duce
	\ifdim\dimen0 = -90\r@dian \then
	   \dimen4 = -1\r@dian
	   \c@mputefalse
	\fi
	\ifdim\dimen0 = 90\r@dian \then
	   \dimen4 = 1\r@dian
	   \c@mputefalse
	\fi
	\ifdim\dimen0 = 0\r@dian \then
	   \dimen4 = 0\r@dian
	   \c@mputefalse
	\fi
	\ifc@mpute \then
		\divide\dimen0 by 180
		\dimen0=3.141592654\dimen0
		\dimen 2 = 3.1415926535897963\r@dian 
		\divide\dimen 2 by 2 
		\Mess@ge {Sin: calculating Sin of \nodimen 0}%
		\count 0 = 1 
		\dimen 2 = 1 \r@dian 
		\dimen 4 = 0 \r@dian 
		\loop
			\ifnum	\dimen 2 = 0 
			\then	\stillc@nvergingfalse 
			\else	\stillc@nvergingtrue
			\fi
			\ifstillc@nverging 
			\then	\term {\count 0} {\dimen 0} {\dimen 2}%
				\advance \count 0 by 2
				\count 2 = \count 0
				\divide \count 2 by 2
				\ifodd	\count 2 
				\then	\advance \dimen 4 by \dimen 2
				\else	\advance \dimen 4 by -\dimen 2
				\fi
		\repeat
	\fi		
			\xdef \sine {\nodimen 4}%
       }}

\def\Cosine#1{\ifx\sine\UnDefined\edef\Savesine{\relax}\else
		             \edef\Savesine{\sine}\fi
	{\dimen0=#1\r@dian\advance\dimen0 by 90\r@dian
	 \Sine{\nodimen 0}
	 \xdef\cosine{\sine}
	 \xdef\sine{\Savesine}}}	      

\def\psdraft{
	\def\@psdraft{0}
}
\def\psfull{
	\def\@psdraft{100}
}

\psfull

\newif\if@scalefirst
\def\psscalefirst{\@scalefirsttrue}
\def\psrotatefirst{\@scalefirstfalse}
\psrotatefirst

\newif\if@draftbox
\def\psnodraftbox{
	\@draftboxfalse
}
\def\psdraftbox{
	\@draftboxtrue
}
\@draftboxtrue

\newif\if@prologfile
\newif\if@postlogfile
\def\pssilent{
	\@noisyfalse
}
\def\psnoisy{
	\@noisytrue
}
\psnoisy
\newif\if@bbllx
\newif\if@bblly
\newif\if@bburx
\newif\if@bbury
\newif\if@height
\newif\if@width
\newif\if@rheight
\newif\if@rwidth
\newif\if@angle
\newif\if@clip
\newif\if@verbose
\def\@p@@sclip#1{\@cliptrue}

\newif\if@decmpr


\def\@p@@sfigure#1{\def\@p@sfile{null}\def\@p@sbbfile{null}
	        \openin1=#1.bb
		\ifeof1\closein1
	        	\openin1=\figurepath#1.bb
			\ifeof1\closein1
			        \openin1=#1
				\ifeof1\closein1%
				       \openin1=\figurepath#1
					\ifeof1
					   \ps@typeout{Error, File #1 not found}
						\if@bbllx\if@bblly
				   		\if@bburx\if@bbury
			      				\def\@p@sfile{#1}%
			      				\def\@p@sbbfile{#1}%
							\@decmprfalse
				  	   	\fi\fi\fi\fi
					\else\closein1
				    		\def\@p@sfile{\figurepath#1}%
				    		\def\@p@sbbfile{\figurepath#1}%
						\@decmprfalse
	                       		\fi%
			 	\else\closein1%
					\def\@p@sfile{#1}
					\def\@p@sbbfile{#1}
					\@decmprfalse
			 	\fi
			\else
				\def\@p@sfile{\figurepath#1}
				\def\@p@sbbfile{\figurepath#1.bb}
				\@decmprtrue
			\fi
		\else
			\def\@p@sfile{#1}
			\def\@p@sbbfile{#1.bb}
			\@decmprtrue
		\fi}

\def\@p@@sfile#1{\@p@@sfigure{#1}}

\def\@p@@sbbllx#1{
		\@bbllxtrue
		\dimen100=#1
		\edef\@p@sbbllx{\number\dimen100}
}
\def\@p@@sbblly#1{
		\@bbllytrue
		\dimen100=#1
		\edef\@p@sbblly{\number\dimen100}
}
\def\@p@@sbburx#1{
		\@bburxtrue
		\dimen100=#1
		\edef\@p@sbburx{\number\dimen100}
}
\def\@p@@sbbury#1{
		\@bburytrue
		\dimen100=#1
		\edef\@p@sbbury{\number\dimen100}
}
\def\@p@@sheight#1{
		\@heighttrue
		\dimen100=#1
   		\edef\@p@sheight{\number\dimen100}
}
\def\@p@@swidth#1{
		\@widthtrue
		\dimen100=#1
		\edef\@p@swidth{\number\dimen100}
}
\def\@p@@srheight#1{
		\@rheighttrue
		\dimen100=#1
		\edef\@p@srheight{\number\dimen100}
}
\def\@p@@srwidth#1{
		\@rwidthtrue
		\dimen100=#1
		\edef\@p@srwidth{\number\dimen100}
}
\def\@p@@sangle#1{
		\@angletrue
		\edef\@p@sangle{#1} 
}
\def\@p@@ssilent#1{ 
		\@verbosefalse
}
\def\@p@@sprolog#1{\@prologfiletrue\def\@prologfileval{#1}}
\def\@p@@spostlog#1{\@postlogfiletrue\def\@postlogfileval{#1}}
\def\@cs@name#1{\csname #1\endcsname}
\def\@setparms#1=#2,{\@cs@name{@p@@s#1}{#2}}
%
%
\def\ps@init@parms{
		\@bbllxfalse \@bbllyfalse
		\@bburxfalse \@bburyfalse
		\@heightfalse \@widthfalse
		\@rheightfalse \@rwidthfalse
		\def\@p@sbbllx{}\def\@p@sbblly{}
		\def\@p@sbburx{}\def\@p@sbbury{}
		\def\@p@sheight{}\def\@p@swidth{}
		\def\@p@srheight{}\def\@p@srwidth{}
		\def\@p@sangle{0}
		\def\@p@sfile{} \def\@p@sbbfile{}
		\def\@p@scost{10}
		\def\@sc{}
		\@prologfilefalse
		\@postlogfilefalse
		\@clipfalse
		\if@noisy
			\@verbosetrue
		\else
			\@verbosefalse
		\fi
}
%
%
\def\parse@ps@parms#1{
	 	\@psdo\@psfiga:=#1\do
		   {\expandafter\@setparms\@psfiga,}}
%
%
\newif\ifno@bb
\def\bb@missing{
	\if@verbose{
		\ps@typeout{psfig: searching \@p@sbbfile \space  for bounding box}
	}\fi
	\no@bbtrue
	\epsf@getbb{\@p@sbbfile}
        \ifno@bb \else \bb@cull\epsf@llx\epsf@lly\epsf@urx\epsf@ury\fi
}	
\def\bb@cull#1#2#3#4{
	\dimen100=#1 bp\edef\@p@sbbllx{\number\dimen100}
	\dimen100=#2 bp\edef\@p@sbblly{\number\dimen100}
	\dimen100=#3 bp\edef\@p@sbburx{\number\dimen100}
	\dimen100=#4 bp\edef\@p@sbbury{\number\dimen100}
	\no@bbfalse
}
\newdimen\p@intvaluex
\newdimen\p@intvaluey
\def\rotate@#1#2{{\dimen0=#1 sp\dimen1=#2 sp
		  \global\p@intvaluex=\cosine\dimen0
		  \dimen3=\sine\dimen1
		  \global\advance\p@intvaluex by -\dimen3
		  \global\p@intvaluey=\sine\dimen0
		  \dimen3=\cosine\dimen1
		  \global\advance\p@intvaluey by \dimen3
		  }}
\def\compute@bb{
		\no@bbfalse
		\if@bbllx \else \no@bbtrue \fi
		\if@bblly \else \no@bbtrue \fi
		\if@bburx \else \no@bbtrue \fi
		\if@bbury \else \no@bbtrue \fi
		\ifno@bb \bb@missing \fi
		\ifno@bb \ps@typeout{FATAL ERROR: no bb supplied or found}
			\no-bb-error
		\fi
		%
%
		\count203=\@p@sbburx
		\count204=\@p@sbbury
		\advance\count203 by -\@p@sbbllx
		\advance\count204 by -\@p@sbblly
		\edef\ps@bbw{\number\count203}
		\edef\ps@bbh{\number\count204}
		\if@angle 
			\Sine{\@p@sangle}\Cosine{\@p@sangle}
	        	{\dimen100=\maxdimen\xdef\r@p@sbbllx{\number\dimen100}
					    \xdef\r@p@sbblly{\number\dimen100}
			                    \xdef\r@p@sbburx{-\number\dimen100}
					    \xdef\r@p@sbbury{-\number\dimen100}}
%
                        \def\minmaxtest{
			   \ifnum\number\p@intvaluex<\r@p@sbbllx
			      \xdef\r@p@sbbllx{\number\p@intvaluex}\fi
			   \ifnum\number\p@intvaluex>\r@p@sbburx
			      \xdef\r@p@sbburx{\number\p@intvaluex}\fi
			   \ifnum\number\p@intvaluey<\r@p@sbblly
			      \xdef\r@p@sbblly{\number\p@intvaluey}\fi
			   \ifnum\number\p@intvaluey>\r@p@sbbury
			      \xdef\r@p@sbbury{\number\p@intvaluey}\fi
			   }
			\rotate@{\@p@sbbllx}{\@p@sbblly}
			\minmaxtest
			\rotate@{\@p@sbbllx}{\@p@sbbury}
			\minmaxtest
			\rotate@{\@p@sbburx}{\@p@sbblly}
			\minmaxtest
			\rotate@{\@p@sbburx}{\@p@sbbury}
			\minmaxtest
			\edef\@p@sbbllx{\r@p@sbbllx}\edef\@p@sbblly{\r@p@sbblly}
			\edef\@p@sbburx{\r@p@sbburx}\edef\@p@sbbury{\r@p@sbbury}
		\fi
		\count203=\@p@sbburx
		\count204=\@p@sbbury
		\advance\count203 by -\@p@sbbllx
		\advance\count204 by -\@p@sbblly
		\edef\@bbw{\number\count203}
		\edef\@bbh{\number\count204}
}
%
%
\def\in@hundreds#1#2#3{\count240=#2 \count241=#3
		     \count100=\count240	
		     \divide\count100 by \count241
		     \count101=\count100
		     \multiply\count101 by \count241
		     \advance\count240 by -\count101
		     \multiply\count240 by 10
		     \count101=\count240	
		     \divide\count101 by \count241
		     \count102=\count101
		     \multiply\count102 by \count241
		     \advance\count240 by -\count102
		     \multiply\count240 by 10
		     \count102=\count240	
		     \divide\count102 by \count241
		     \count200=#1\count205=0
		     \count201=\count200
			\multiply\count201 by \count100
		 	\advance\count205 by \count201
		     \count201=\count200
			\divide\count201 by 10
			\multiply\count201 by \count101
			\advance\count205 by \count201
		     \count201=\count200
			\divide\count201 by 100
			\multiply\count201 by \count102
			\advance\count205 by \count201
		     \edef\@result{\number\count205}
}
\def\compute@wfromh{
		\in@hundreds{\@p@sheight}{\@bbw}{\@bbh}
		\edef\@p@swidth{\@result}
}
\def\compute@hfromw{
	        \in@hundreds{\@p@swidth}{\@bbh}{\@bbw}
		\edef\@p@sheight{\@result}
}
\def\compute@handw{
		\if@height 
			\if@width
			\else
				\compute@wfromh
			\fi
		\else 
			\if@width
				\compute@hfromw
			\else
				\edef\@p@sheight{\@bbh}
				\edef\@p@swidth{\@bbw}
			\fi
		\fi
}
\def\compute@resv{
		\if@rheight \else \edef\@p@srheight{\@p@sheight} \fi
		\if@rwidth \else \edef\@p@srwidth{\@p@swidth} \fi
}
%
\def\compute@sizes{
	\compute@bb
	\if@scalefirst\if@angle
	\if@width
	   \in@hundreds{\@p@swidth}{\@bbw}{\ps@bbw}
	   \edef\@p@swidth{\@result}
	\fi
	\if@height
	   \in@hundreds{\@p@sheight}{\@bbh}{\ps@bbh}
	   \edef\@p@sheight{\@result}
	\fi
	\fi\fi
	\compute@handw
	\compute@resv}

%
%
\def\psfig#1{\vbox {
	%
	\ps@init@parms
	\parse@ps@parms{#1}
	\compute@sizes
	\ifnum\@p@scost<\@psdraft{
		\special{ps::[begin] 	\@p@swidth \space \@p@sheight \space
				\@p@sbbllx \space \@p@sbblly \space
				\@p@sbburx \space \@p@sbbury \space
				startTexFig \space }
		\if@angle
			\special {ps:: \@p@sangle \space rotate \space} 
		\fi
		\if@clip{
			\if@verbose{
				\ps@typeout{(clip)}
			}\fi
			\special{ps:: doclip \space }
		}\fi
		\if@prologfile
		    \special{ps: plotfile \@prologfileval \space } \fi
		\if@decmpr{
			\if@verbose{
				\ps@typeout{psfig: including \@p@sfile.Z \space }
			}\fi
			\special{ps: plotfile "`zcat \@p@sfile.Z" \space }
		}\else{
			\if@verbose{
				\ps@typeout{psfig: including \@p@sfile \space }
			}\fi
			\special{ps: plotfile \@p@sfile \space }
		}\fi
		\if@postlogfile
		    \special{ps: plotfile \@postlogfileval \space } \fi
		\special{ps::[end] endTexFig \space }
		\vbox to \@p@srheight sp{
			\hbox to \@p@srwidth sp{
				\hss
			}
		\vss
		}
	}\else{
		\if@draftbox{		
			\hbox{\frame{\vbox to \@p@srheight sp{
			\vss
			\hbox to \@p@srwidth sp{ \hss \@p@sfile \hss }
			\vss
			}}}
		}\else{
			\vbox to \@p@srheight sp{
			\vss
			\hbox to \@p@srwidth sp{\hss}
			\vss
			}
		}\fi

	}\fi
}}
\psfigRestoreAt
\let\@=\LaTeXAtSign

\title{Depinning of a domain wall in the 2d random-field Ising model}
\author{Barbara Drossel$^1$ and Karin Dahmen$^2$}
\address{${}^1$ Department of Theoretical Physics, 
  University of Manchester, 
  Manchester M13 9PL, England} 
\address{${}^2$ Lyman Laboratory of Physics, 
  Harvard University, 
  Cambridge, MA02138, USA} 
\date{\today}
\maketitle
\begin{abstract}
We report studies of the behaviour of a single driven domain wall in the
2-dimensional non-equilibrium zero temperature random-field Ising model,
closely above the depinning threshold. It is found that even for very
weak disorder, the domain wall moves through the system in percolative 
fashion. At depinning, the fraction of spins that are flipped by the 
proceeding avalanche vanishes with the same exponent $\beta=5/36$
as the infinite percolation cluster in percolation theory. 
With decreasing disorder strength, however, the size of the critical 
region decreases. Our
numerical simulation data appear to reflect a crossover behaviour
to an exponent $\beta'=0$ at zero disorder strength. 
The conclusions of this paper strongly rely on analytical arguments. 
A scaling theory in
terms of the disorder strength and the magnetic field is presented
that gives the values of all critical exponent except for one, the
value of which is estimated from scaling arguments.
\end{abstract}
\begin{multicols}{2}
\section{Introduction}
 The random-field Lenz-Ising model is one of the simplest examples
of random media, with applications far beyond magnetic systems.  (For
a recent review, see \cite{nat97}.) Recently interesting
nonequilibrium aspects have been studied, such as the field driven
motion of a single domain wall
\cite{ji91,mar91,nat92,nar93,les97,ji92,koi92,koi92a}, domain
coarsening in rapidly cooled magnetic systems, where the domain wall
motion is curvature driven \cite{aep86}, and hysteresis, with
many interacting driven domain walls\cite{per97}.  Studies of a single
driven domain wall were found to describe fluid invasion in
porous media \cite{ji91}.  These studies fall into the large class of
interface depinning problems, which also include charge density wave
depinning, contact line depinning, earthquakes, and domain wall
depinning in magnets\cite{depinning,Zap97}.  In these systems, second
order dynamical phase transitions were found as the driving force
$F$ surpasses some critical threshold value $F_c$, at which the
interface becomes depinned, and starts to propagate through the system
at a velocity $v \sim (F-F_c)^\phi$ with a critical exponent $\phi$.

In previous work on a single field driven domain wall in the random
field Ising model in three dimensions, three different modes of
interface propagation close to the depinning threshold were identified
\cite{ji92}: For weak, bounded disorder, the marginally stable
interface at $F_c$ is facetted, for intermediate disorder it is
self-affine, and for large disorder it is self-similar.  In the
``faceted growth'' \cite{mar91,ji91,ji92,koi92a} the interface
propagates just as in the absence of disorder, penetrating the medium
completely, with a roughness exponent $\zeta=0$ at the depinning
threshold. This type of interface motion can occur in any dimension,
but only for a narrow, bounded distribution of random fields, not for
unbounded (Gaussian) distributions of random fields.  The existence
of this phase is lattice dependent \cite{koi92}.

In the self-affine regime, which was seen in 3-dimensional
simulations with bounded and Gaussian distribution of random fields
\cite{ji92,Robbins},
neighbouring interface segments proceed
coherently, and the interface has a roughness exponent smaller than
one. Overhangs and ``bubbles'' (i.e., uninvaded domains left behind by
the advancing interface) occur only below a certain length scale and
can therefore be neglected on long length scales, where the interface can
be described by a single-valued function. The critical properties of
the interface near the depinning threshold were derived analytically
starting from a continuum model with a single-valued function for the
interface, and performing a renormalization group calculation, and 
$\epsilon$ expansion around the upper critical dimension which is 5
\cite{nat92,nar93}.  This $\epsilon$ expansion yields the roughness
exponent $\zeta=(5-d)/3$ for a $d-1$-dimensional domain wall in a in
a $d$-dimensional system, which is argued to be exact to all orders
in perturbation theory \cite{nar93}, however, numerical simulations
show deviations from this prediction \cite{les97}. If the
renormalization group result $\zeta=(5-d)/3$ is indeed exact, it
implies that $d=2$ is the lower critical dimension, where the ansatz
of an interface without overhangs at large scales breaks down and
conventional correlated depinning does not occur any more. 
Anisotropies in the medium may give rise to further depinning 
universality classes\cite{tan95}.

For strong disorder, the invading phase advances in a
percolation-like manner, following routes of particularly high random
field values. When the driving force is at the depinning threshold,
the invading phase penetrates only a vanishing volume fraction of the
invaded medium, just as a spanning cluster in percolation theory
\cite{ji91,ji92}.  Numerical results for the fractal 
dimension of the invaded volume and the external hull of the interface 
suggest that
this system is in the same universality class as uncorrelated site
percolation \cite{ji91,ji92}.  When the disorder strength is
decreased, the percolation pattern coarsens, and the thickness
of the percolation fingers increases and diverges at the critical
disorder $R_c$ which marks a transition to
conventional (coherent, self-affine) 
depinning \cite{ji91,ji92,koi92}. While in three dimensions $R_c > 0$,
in 2 dimensions simulation
results seem to indicate that this divergence of the finger width 
occurs only in the limit of zero disorder ($R_c=0$)\cite{ji91,koi92},
suggesting again that d=2 plays the role of a lower critical
dimension.

This result however so far is only an indication. Within the numerical
accuracy of the 2-dimensional simulations a nonzero $R_c$ could 
not definitely be ruled out, and the issue is still controversial.
While 2 is the lower critical dimension for the equilibrium random-field 
Ising model, there is no obvious reason that this result should be 
tranferable to a nonequilibrium situation \cite{footnote2}.
There are other nonequilibrium problems with similar hurdles
to establishing the lower critical dimension. An example 
are hysteresis loops in the 2-dimensional random-field Ising 
model with many interacting interfaces. Numerical simulations,
even of rather large systems (up to $30000^2$), seem to converge
towards zero critical disorder, but so far do not definitely rule 
out the possibility of a phase transition at nonvanishing disorder 
value either\cite{per97}.  

Another open question is whether the type of correlated percolation 
found for domain-wall motion in the random-field Ising model does
indeed belong to the same universality class as conventional uncorrelated 
site percolation. In fact, while the fractal dimension of the external
perimeter of a site percolation cluster is 4/3 \cite{sta94}, in agreement
with the dimension found for the magnetic interface \cite{ji91}, the
fractal dimension of the hull (which is the perimeter that one
measures stepping along occupied sites) of a percolation cluster is
7/4 \cite{sta94}, which is different from the value 4/3 for the
RFIM interface (see below). 

The main purpose of this paper is to present new support for the
conclusion that 2 is indeed the lower critical dimension for the
transition described above, with $R_c=0$, and to give analytical and
numerical arguments that in 2 dimensions on sufficiently long
length scales (even for very narrow Gaussian distribution of random
fields) the interface propagates in the self-similar mode,
characterized by site percolation critical exponents.  In contrast to
previous simulations of the pinned interface where the 
depinning point is approached from below, here it is approached 
from {\it above} with focus on the volume fraction $m$ filled by the 
invading phase as a
function of the strength of the driving force (i.e., the magnetic
field). Since $m$ vanishes at percolation-like depinning, but is
nonzero at correlated depinning, it can be used as an ``order parameter''
for the phase transition between the two behaviours.  
Scaling forms for $m$, and the width
and length and fractal dimension of the interface, and other
quantities, are conjectured and tested numerically, and ultimately
justified by a scaling theory.  Analytical arguments based on the
properties of the domains of unflipped spins left untouched by the
proceeding interface give strong support for a phase transition at
zero disorder. They also show that the width of the critical region
decreases towards zero as the strength of the disorder vanishes.
Mappings of the model for several values of the parameters onto other
percolation models with (supposedly) known values for the critical
exponents confirm that the growth of $m$ above depinning is
characterized by the conventional percolation exponent $\beta=5/36$.
The numerical data are compatible with a critical
point at zero random field, and reflect a crossover from the percolation
critical exponent $\beta=5/36$ very close to depinning to $\beta'=0$ 
further away. The roughness exponent characterizing the interface and the
fractal dimension of the interface are $\zeta=1$ and $d_f= 4/3$.
Building on these results, we conjecture a scaling theory for the
critical behaviour in $d=2$ dimensions in the limit of small
disorder. Except for one exponent, the exact values of all other
exponents can be postulated, and for the remaining exponent, an
approximate result is obtained from scaling arguments.

The outline of the remainder of the paper is as follows: In
section \ref{model}, the model is introduced. In sections
\ref{numerical} and \ref{analytical}, the numerical and analytical
results are given. The final section contains a discussion of the
results and ideas for further work.

\section{The model} \label{model}
The random-field Ising model is defined by the following Hamiltonian
\begin{equation}
{\cal H} = -\sum_{\langle i,j \rangle}S_iS_j-\sum_i(h_i+H)S_i\, .
\end{equation}
The field $H$ is the external field, $\langle i,j \rangle$ denotes 
nearest-neighbour pairs, and the spin variable $S_i$ assumes the values 
$\pm 1$. 
In most discussions of the model, the probability distribution of 
the random fields $\{h_i\}$ is a Gaussian
\begin{equation}
p(h_i) = {\exp{\left[-h_i^2/2R^2\right]} \over \sqrt{2\pi}R}
\end{equation}
of width $R$.
Random fields at different sites are taken to be uncorrelated.

The subsequent discussion is limited to a 2-dimensional system on a
square lattice at zero temperature. 
Initially all spins are taken to be 
pointing down ($S_i=-1$), except for one column of up-spins at the
left boundary of the system, which, for convenience are given 
an infinitely positive random field $h_i$. 
The upper and lower sides of the system are
connected by periodic boundary conditions.
For a given value $H$ of the 
external magnetic field, successively all those spins are flipped up
that have a positive local field $J\sum_j S_j + H +
h_i$ and at least one flipped ``up'' neighbour ({\it i.e.} those spins
immediately neighbouring the successively propagating interface of the 
growing cluster of up spins). 
For negative or small positive values of the
external field $H$, the interface of up spins 
cannot proceed far and
stops after a small number of steps. For large positive $H$, on the
other hand, the propagating interfaces runs from left to right through
the entire system, leaving behind only small domains of unflipped
spins. For values $H$ below some critical value $H_c(R)$, the order
parameter $m$ vanishes, i.e., $m=0$, while it becomes finite for 
$H > H_c(R)$. If at $R>0$ the spin-flip avalanche proceeds percolation-like,
the transition from $m=0$ to $m>0$ is continuous with
\begin{equation}
m \simeq A(R) (H-H_c(R))^\beta\, , \label{beta}
\end{equation}
with some prefactor $A$ that depends
on the width of the random field
distribution.  If the invaded area has the same critical properties as
a spanning cluster in uncorrelated site percolation (in 2 dimensions), 
as suggested in
\cite{ji91,koi92}, then $\beta = \beta_{\text{perc}}= 5/36 \simeq
0.139$.  If 2 is the lower critical dimension for the 
transition from percolation-like to correlated depinning, we expect that
this percolation-like behaviour persists even for arbitrarily small 
randomness $R$ on sufficiently long length scales and at sufficiently
small magnetic fields. On the
other hand, if there is a phase transition to correlated depinning for
some finite random field strength $R_c$, then the magnetization
is expected to display a jump from $0$ to some finite value $m$ at
at $H=H_c(R)$ for $R<R_c$. 
In this paper, we will argue for $R_c =0 $ and the scenario 
of equation (\ref{beta}) for $R > 0$ and $H$ sufficiently close to 
depinning $H_c(R)$ on sufficiently long length scales.
If $$H+h_i+(2n-4)J >0 > H+h_i+(2(n-1)-4)J\,,$$ the local field at site
$i$ becomes positive (causing $S_i$ to flip up)
when the $n$th nearest neighbour flips up. It is
useful to define the probabilities 
\begin{equation} \rho_n=
\int\limits_{(4-2n)J-H}^{(6-2n)J-H} p(h_i)\text{d}h_i\,  \label{rho}
\end{equation} 
 that a spin flips as soon as $n$ of its nearest neighbours are
flipped. Since we only allow spins connected to the 
advancing interface to flip,
({\it i.e.} isolated spins remain unflipped), we absorb
$\rho_0$ into $\rho_1$. For the advancement of the avalanche, there is
no difference between a site that flips only when all four neighbours
are flipped, and a site that does not flip even with four flipped
neighbours, so we include in $\rho_4$ all sites with $h_i<-H-2J$. 
Subsequently, for convenience, we describe the
system in terms of these four probabilities instead of $H/J$ and
$R/J$, as done also by other authors\cite{footnote}. 
They span a three dimensional parameter
space, since $\rho_1+\rho_2+\rho_3+\rho_4=1$. 
The plane spanned by $H/J$ and $R/J$ represents a cut through this 
space.
Changing the external field $H$ for a given distribution of random
fields corresponds to moving 
along a line given implicitely by Eq.~(\ref{rho}). There exists a
critical surface in the parameter space that separates the region with
$m=0$ from the region with $m>0$, which
contains the line $H_c(R)$.

We begin by presenting some fundamental properties of the two limiting
cases of weak and strong disorder.
Figure~\ref{limits} illustrates the relation between the
model parameters and the densities $\rho_n$.
For small $R/J$ (weak disorder), regions of a given
$n$ are wide compared to $R/J$, and one or two neighbouring values of
$n$ dominate the system, while for large $R/J$ (strong disorder), 
the regions of given $n$ are
narrow, and the two boundary regions for $n=1$ and $n=4$
dominate. Obviously, in the limit $R/J\to \infty$, the model
corresponds to a site percolation system, where $\rho_4=1-\rho_1$.
Knowing the value of the
site percolation threshold, $\rho_1^c\simeq 0.59$, we can immediately
give an implicit expression for the critical magnetic field $H_c$,
$$\rho_1^c=\int\limits_{2J-H_c}^\infty p(h_i)\text{d}h_i,\qquad 
R/J \rightarrow \infty. $$
\begin{figure} 
\narrowtext
\centerline{\psfig{file=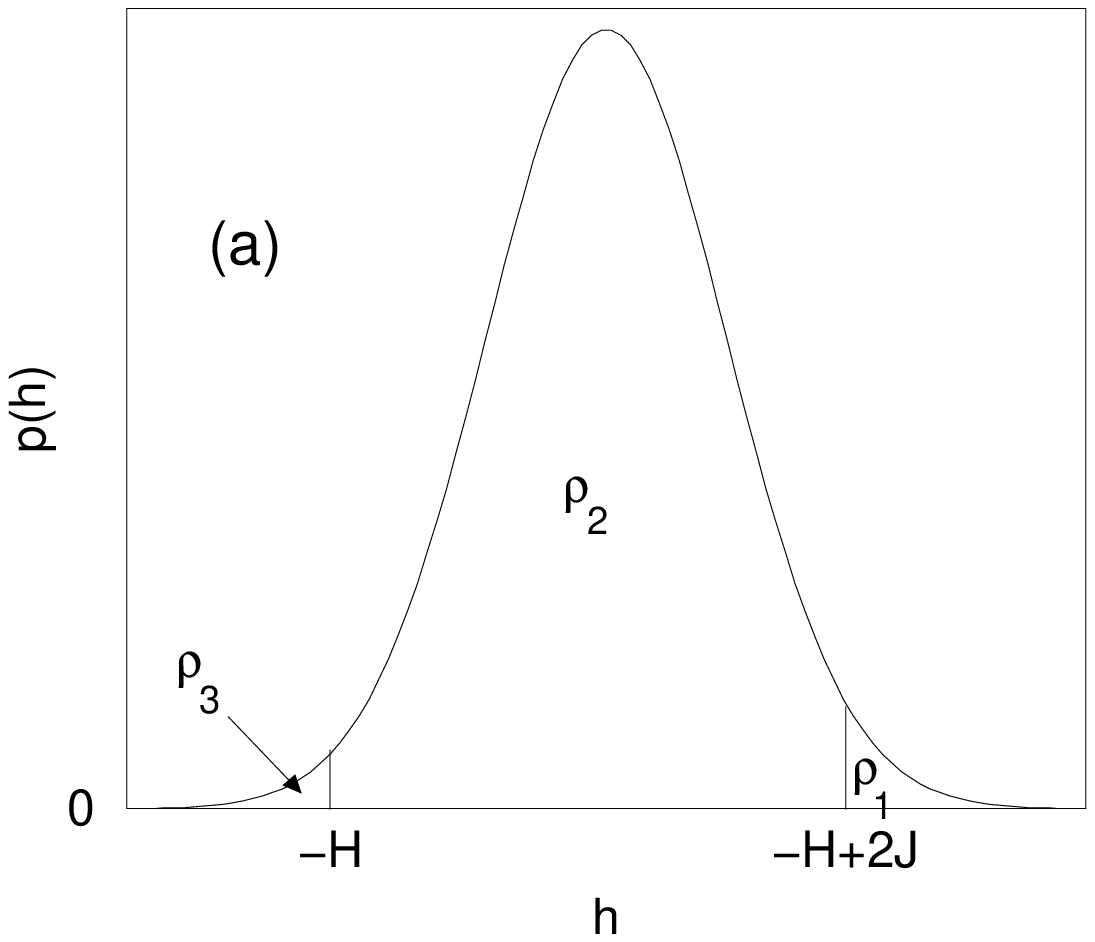,height=2.5in,angle=-90}}
\centerline{\psfig{file=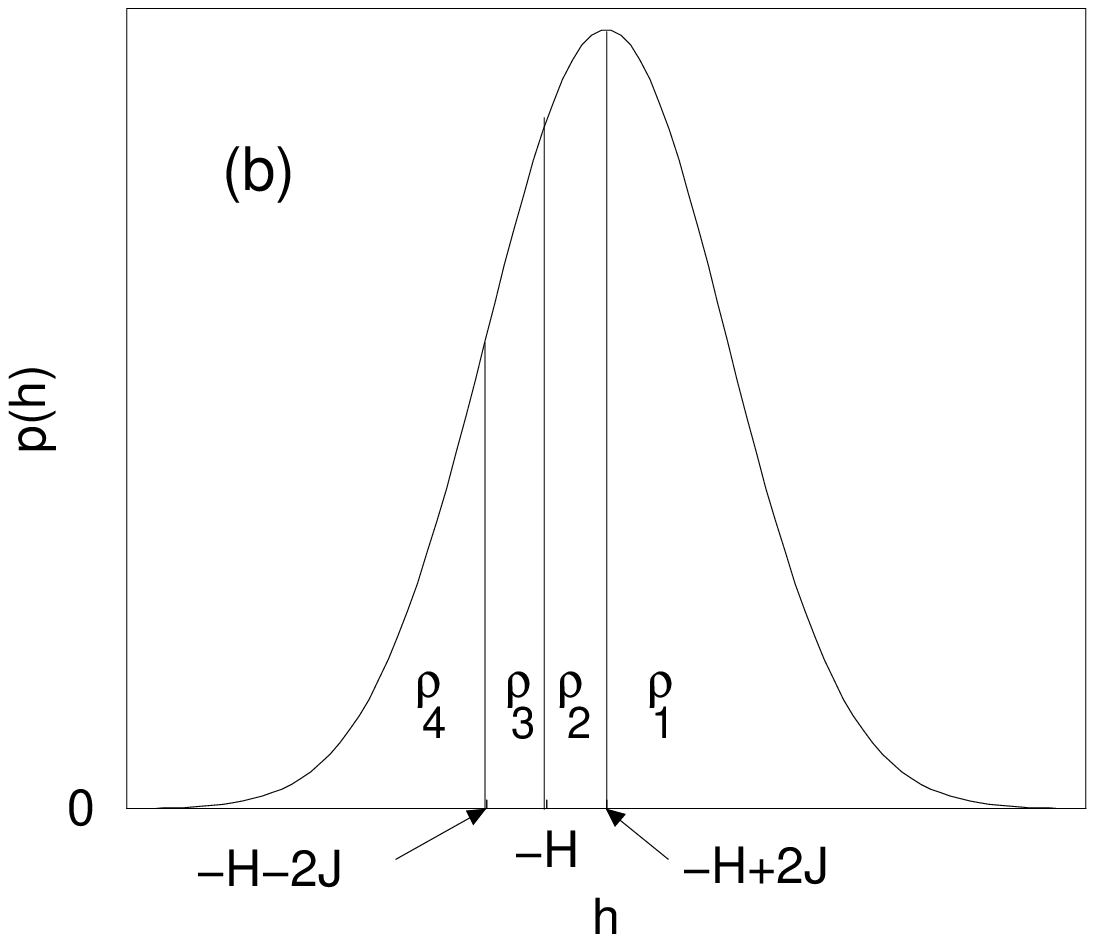,height=2.5in,angle=-90}}
\caption{The densities $\rho_n$ for (a) small $R/J$ and (b) large $R/J$, 
represented as areas under the Gauss-distribution for the random field. 
In (a), the density $\rho_4$ is so small, and its area is so far to the left,
that it is not shown in the figure.}
\label{limits}
\end{figure}

In the limit of small $R/J$, depinning occurs when $H$ is such that
$\rho_2$ is close to one, and $\rho_1 \simeq \sqrt{\rho_3}$. To
understand this, consider a stable phase boundary between the spin up
and spin down regions, as drawn in figure~\ref{boundary}.  The invaded
area (spin up) is indicated in black, as well as those spins that will
flip as soon as one of their neighbours is flipped. The grey sites
will only flip when three neighbours are flipped and sit therefore in
the corners of the boundary. White sites will flip when two neighbours
are flipped. The depinning transition occurs when there exists no
stable boundary that spans the system. Let us call a possible boundary
"pinning path".  The following construction of a pinning path gives a
good estimate for the relation between $\rho_1$ and $\rho_3$ at the
depinning threshold: Consider a system with all spins down. Now start
at the bottom of the system at a bond that has no black right-hand
neighbour and make a step upward. The path can proceed in the same
direction as long as there are no black right-hand neighbours. It can
turn left anywhere (if this does not lead to a black right-hand
neighbour), but can turn right only at grey sites. Since the path is
not allowed to intersect itself, and since it must ultimately arrive
at the top end of the system, the mean number of right turns must
equal the mean number of left turns. Thus, for each black site that is
avoided by a left turn, there must be a grey site, where a right turn
can be made. Since the density of black sites is $\rho_1$, the path
has on an average $1/\rho_1$ opportunities to turn left before
encountering the next black site. After the left turn, the probability
of encountering a grey site before encountering a black site, is
\begin{equation}
p_{surv}=\rho_3/(\rho_3+\rho_1)\, .\label{survival}
\end{equation}
\begin{figure} \narrowtext
\centerline{\psfig{file=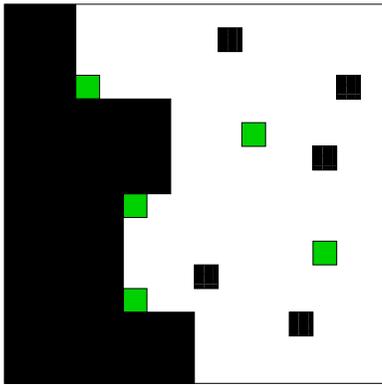,height=2in,angle=-90}}
\caption{The phase boundary between up spins (large black area) and down spins 
(remaining area). The down spins ar colour coded depending on whether they flip 
as soon as one (black), two (white), or three neighbours (grey) are flipped.}
\label{boundary}
\end{figure}
 The path can survive when this probability, multiplied by the number
of turning opportunities, is not smaller than 1, i.e.,
$\rho_3/\rho_1(\rho_3+\rho_1) \ge 1$, leading to $$ \rho_3 =
\rho_1^2/(1-\rho_1)$$ at the depinning threshold. For small values of
$R$ (weak random fields), $\rho_3$ and $\rho_1$ are also small, and
$\rho_3 \simeq \rho_1^2$ at the depinning threshold. In \cite{koi92},
the relation $\rho_3 \propto \rho_1^{1.75\pm 0.05}$ was obtained
numerically (for a rectangular distribution of random fields), 
however, an exponent 2 can also be reconciled with their
figure~4, when the smaller slope in the lower part of the plot is
ascribed to finite-size effects. In \cite{blo97}, the relation
$\rho_3 \simeq (2/3)\rho_1^2$ was derived, which agrees with the one
given here apart from the prefactor. The authors of \cite{blo97}
obtained their relation from the condition that an initially straight
interface along the first column of sites can invade the same number
of sites in the second column of a 2-column system as in the third
column of a three-column system. The prefactor should therefore change
when more columns are taken into account.  The argument presented in
this paper does not consider the possibility of having two right (or
left) turns in sequence, which, however, occurs by a factor
$\rho_3/(\rho_1+\rho_3)$ less often than alternating turns and makes
therefore a negligible contribution to the above calculation in the
limit $\rho_1 \to 0$. The argument also neglects possible correlations
between different pinning paths starting at the same initial
point. Taking these into account will probably change the prefactor.
Irrespective of the prefactor, however,
we can easily see that the density $\rho_4$ is negligible in the
limit of small disorder: A short calculation gives the approximate
result $\lim_{R\to 0}H_c(R) \simeq 2  (2-\sqrt{2})J \simeq 1.172 J $
(also derived in \cite{blo97}),
leading to $$\rho_3 \simeq (R/2.94 J)
\exp[-1.373J^2/2R^2]$$ and $$\rho_4 \simeq (R/7.95 J)
\exp[-10.06J^2/2R^2]$$ at depinning threshold. 
Thus, the ratio
$\rho_4/\rho_3^4$ becomes arbitrarily small for small disorder, which
means that clusters of four grey sites forming a $2\times 2$-square
occur far more often than isolated sites that do not flip with three
flipped neighbours. Since both play the same role in the system by
blocking an avalanche even when surrounded by it on three sides, and 
since their size difference is irrelevant on long length scales, the
neglection of $\rho_4$ does not change any properties of the system in
the limit of small disorder. 

To conclude this section, let us note
that in addition to pinning paths that span the system and separate
the invaded from the non invaded area, there exist pinning paths that
are closed loops, delimiting unflipped domains within
the invaded region.

\section{Numerical results} \label{numerical} 
Earlier numerical studies\cite{ji91,koi92} used a
quadratic system of $L^2$ sites in which
the magnetic field $H$ was increased incrementally, 
allowing the system to relax after each (adiabatic) increase.
The system size dependent critical field $H_c(R,L)$ at which 
the invading spin-up phase first reached the right boundary of the system,
was seen to converge towards a constant value $H_c(R)$
as $L$ was increased.
Information about the critical field,
the fractal dimension of the invading cluster, and its perimeter
were obtained upon approaching the depinning threshold from
below.

The present work, in contrast, is mainly concerned with the behaviour of
the fraction $m$ of flipped spins {\it above} the depinning threshold
$H > H_c(R)$, focussing on the question whether 
$m$ goes to zero continuously or
discontinuously at depinning, and on the value of the critical
exponent $\beta$. For this purpose, the external magnetic field (or,
equivalently, the densities $\rho_n$) was set to a fixed value
throughout a simulation run, and the interface was allowed to advance 
in a system
of height $L$ until it either came to a halt, or until it reached a
cutoff distance which we chose to be ca. 10.5 $L$. 
Memory was allocated dynamically, and the
system was updated by flipping all spins with positive local field
along the interface.
The following quantities were measured for various values of the
system height $L$, and averaged over up to 250 realizations of the
disorder: (i) the position
of the most advanced and most retarded site, and 
the mean position of the interface, as well as the
interface length at the moment where it came to a stop (if it did so
before running over the maximum allowed distance). (ii) The fraction
of spins flipped by the avalanche (disregarding the first $L/2$
columns, where the interface had not yet reached its stationary
behaviour, and the last columns that were only partially
invaded by the interface). (iii) The size distribution of the patches
of unflipped spins left behind by the advancing interface. Before
evaluation, these patches were allowed to relax, which is realistic
for a magnetic system, but not for fluid invasion in a porous medium,
where trapped regions cannot shrink.  
To study the low disorder regime, in our simulations we set $\rho_4$
to zero and $\rho_3$ to some small fixed value
$\rho_3= 0.2$, 0.1, or 0.05. 
$\rho_1$ was chosen close to the depinning threshold $\rho_1^c(\rho_3)$.
The snapshots in
figure~\ref{snapshots} show two pinned invasion patterns
for two different values of the disorder ($\rho_3$),
at $\rho_1$ slightly below the 
respective depinning threshold $\rho_1^c(\rho_3)$.
\begin{figure} \narrowtext
\centerline{\psfig{file=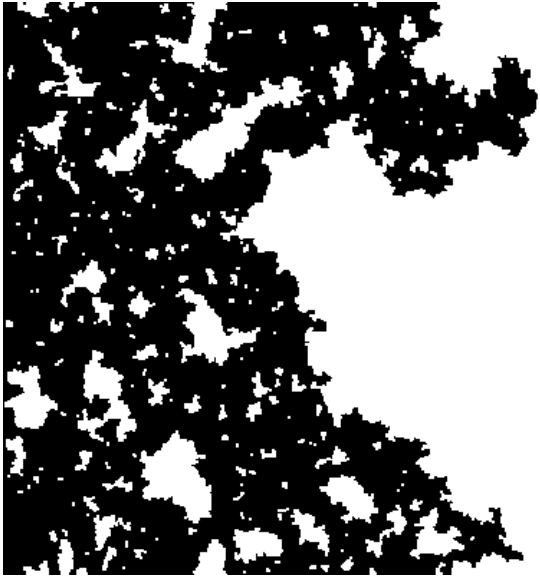,height=3in,angle=-90}}
\vskip 1cm
\centerline{\psfig{file=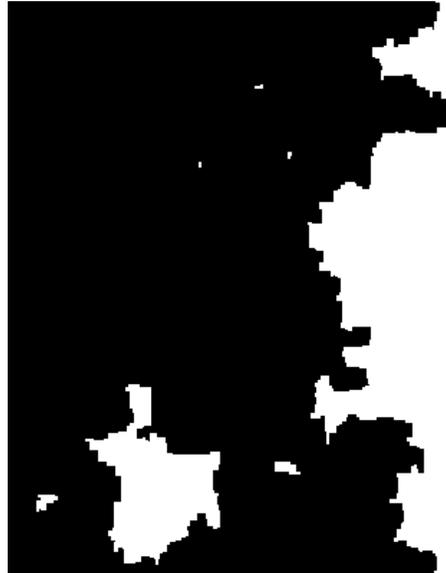,height=3in,angle=-90}}
\caption{The invaded area (black) for $L=400$ in the low disorder 
regime ($\rho_4=0$), slightly below depinning, with
(a) $\rho_3=0.2$ and
$\rho_1=0.43 < \rho_1^c(0.2) \simeq 0.4345$ and (b) $\rho_3=0.05$ and 
$\rho_1=0.2236 < \rho_1^c(0.05) \simeq 0.226$. $L$ is the
linear system size in the vertical direction. Roughly speaking,
in the low disorder regime,
smaller $\rho_3$ corresponds to smaller
disorder, and $(\rho_1-\rho_1^c)$ corresponds to $(H-H_c(R))$.}
\label{snapshots}
\end{figure}

Two characteristic trends can be discerned for decreasing $\rho_3$
(i.e.~ decreasing random field strength $R$): The length of straight
front segments increases, and the number of unflipped domains in the
invaded area decreases. The first feature was explained in the
previous, and the second feature will be explained in the following
section. First, however, let us give more details of the simulation
results.

\subsection{Properties of the interface ($\zeta=1$ and $d_f=4/3$),
and determination of the critical value
$\rho_1^c(\rho_3)$ in the low disorder regime} \label{numinterface} 

Figures~\ref{max} and \ref{min}
show the probability density $p(x_{\text{max}}, L)$
for the position $x_{\text{max}}$ and $x_{\text{min}}$ 
of the most advanced and most retarded site of the interface for
$\rho_3=0.05$ and $\rho_1=0.226 \simeq \rho_1^c(0.05)$, 
scaled by the system height $L$.
At depinning one expects the scaling behaviour
$p(x_{\text{max}}, L) \sim f_1(x_{\text{max}}/L^\zeta)$
with a universal roughness exponent $\zeta$
and a universal scaling function $f_1$. Analogously,
$p(x_{\text{min}}, L) \sim f_2(x_{\text{min}}/L^\zeta)$
with a universal scaling function $f_2$.
The curves for different values of $L$ collapse nicely, indicating
that the system is indeed at the depinning threshold, i.e.,
$\rho_1=\rho_1^c(\rho_3)$, and $\zeta =1 $. 

\begin{figure} \narrowtext
\centerline{\psfig{file=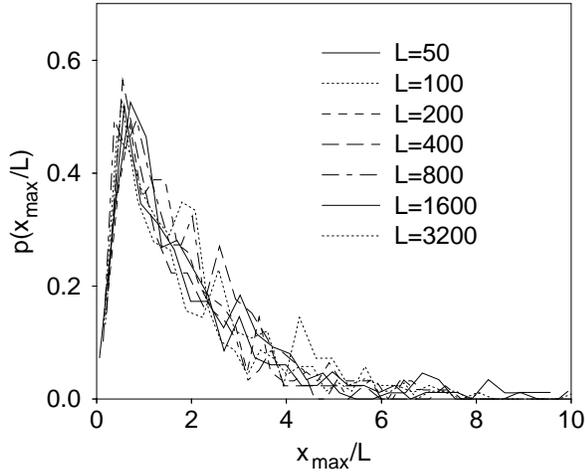,height=2.5in,angle=-90}}
\caption{Probability density for the position of the most advanced
site of the pinned interface, for $\rho_3=0.05$
and $\rho_1=0.226 \simeq \rho_1^c(0.05)$. 
The collapse of the curves indicates that $\zeta=1$.  
\label{max}}
\end{figure}
\begin{figure} \narrowtext
\centerline{\psfig{file=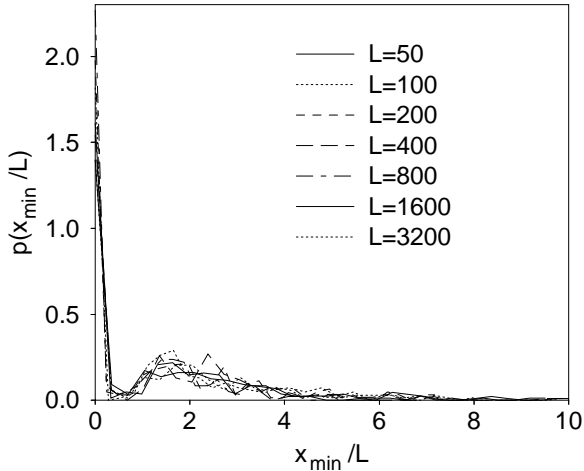,height=2.5in,angle=-90}}
\caption{Probability density for the position of the most retarded site 
of the pinned interface, for $\rho_3=0.05$ and 
$\rho_1=0.226 \simeq \rho_1^c(0.05)$. 
The collapse indicates that $\zeta=1$.
}
\label{min}
\end{figure}
For larger values of $\rho_1$, ({\it i.e.} $H > H_c(R)$)
the peak of
the curves moves to the right with increasing $L$, and a nonvanishing
fraction of all interfaces do not get stuck before reaching the cutoff
distance $10.5L$ (especially for large values of $L$),
since the system is above the
depinning threshold.
For smaller values of
$\rho_1$, {\it i.e.} below the depinning threshold,
the percentage of interfaces that remain attached to the
first column increases with increasing $L$. 
Other values of $\rho_3$ 
give similar results. 
At depinning, one expects 
the mean thickness of the interface, defined as the number of columns
spanned by the interface after it got stuck, to also scale as $L^\zeta$.
For the parameter values of figures \ref{min} and \ref{max}
this was verified with $\zeta =1$.
Once the interface has reached this mean asymptotic thickness
of the order $L$, it becomes pinned with equal probability at
any moment. This is reflected by the exponential tails of the scaling
functions in figures~\ref{max} and \ref{min}.

Figure~\ref{length} shows the mean length $l$ of the pinned interface
as function of $L$ for different values of $\rho_1$, at fixed
$\rho_3=0.1$. This length is the number of flipped spins that
constitute the nearest-neighbour connected interface, {\it i.e.} the
number of sites of the pinning path described in section \ref{model}.
Only interfaces with $x_{\text{min}} > L/2$ and $x_{\text{max}} < 10.5 L $ 
were considered.  The expected scaling form is $l \sim L^{d_f}
g((\rho_1-\rho_{1}^{c}(\rho_3)) L^{1/\nu})$ with universal scaling
function $g$, correlation length exponent $\nu$, and fractal dimension
$d_f$.  At the critical threshold $\rho_1 \simeq 0.317 \simeq
\rho_1^c(0.1)$ the points do indeed follow a power law. For nearby
values of $\rho_1$, the critical behaviour is only visible for $L <
\xi $ where $\xi \sim ((\rho_1-\rho_{1}^{c}(\rho_3))/\rho_1)^{-\nu}$
is the correlation length.  For larger $L$ it crosses over to a
different behaviour (linear in $L$ in the case of $\rho_1>\rho_1^c$). The
straight line in the figure is a power-law fit to the critical data
set, with exponent $d_f = 1.332 \pm 0.007$.  Simulations for other
values of $\rho_3$ give the same exponent $d_f=1/\nu=4/3$. 
For a percolation cluster,
the exponent $d_f$ is $7/4$ \cite{sta94}.  The exponent 4/3 is retrieved,
however, for a percolation cluster hull, when one allows steps to the
next-nearest neighbours as well, thereby bridging the most narrow
throats \cite{sta94}. Since in our model sites with more flipped
neighbours are more likely to flip also, narrow throats will be
bridged with a certain probability, thus producing the exponent 4/3.
The same exponent $d_f=4/3$ characterizes also the fractal dimension of
a self-avoiding random walk. This analogy becomes apparent in the
limit of weak randomness (small $\rho_3$), where interfaces can be
constructed by finding pinning paths that connect "grey" sites, as 
discussed in section \ref{model} above. 
On sufficiently large scales, such paths are essentially
self-avoiding random walks. 

\begin{figure} \narrowtext
\centerline{\psfig{file=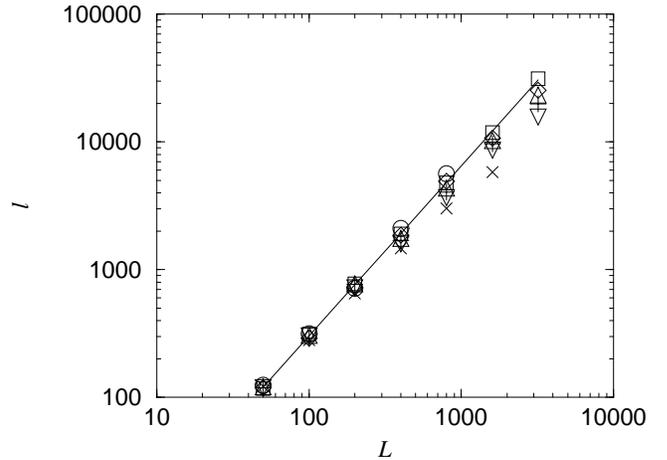,height=2.5in,angle=-90}}
\caption{
The length of the interface $l$ as function of $L$,
for $\rho_3=0.1$ and $\rho_1=$0.315 (circle), 0.317 (square), 0.318
(diamond), 0.319 (triangle up), 0.32 (plus), 0.322 (triangle down),
and 0.325 (x). The straight line is a power law fit to the
$\rho_1=0.317 \simeq \rho_1^c(0.1)$ data, with an exponent $1.332$.
}
\label{length} 
\end{figure}

Figure~\ref{lengthcoll} shows the collapsed data of
Figure~\ref{length}.  On the horizontal axis, the scaling variable $x
\equiv (\rho_1-\rho_{1}^{c}(\rho_3)) L^{4/3}$ is plotted, and on the
vertical axis $l/L^{4/3}$, which is expected to be identical to the
scaling function $g(x)$ defined above. One can see that the scaling
function is constant for small $x$, and decays with $x^{-1/4}$ for
large $x$. This decay corresponds to a linear dependence of $l$ on
$L$. The data points to the largest value of $\rho_1$ are already
outside the scaling regime. The scattering of the other points is due
to not too good statistics. 

\begin{figure} \narrowtext
\centerline{\psfig{file=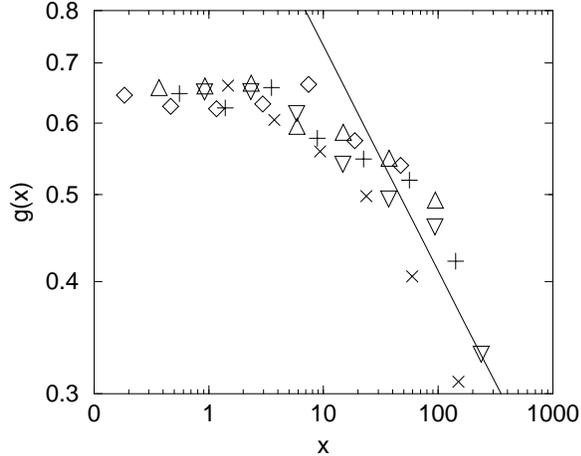,height=2.5in,angle=-90}}
\caption{ The scaling function $g(x)$ for the interface length.  The
symbols are the same as in the previous figure. The straight line is a
power law with exponent $-1/4$.}
\label{lengthcoll} 
\end{figure}
\subsection{The exponent $\beta$} \label{OP}
Figure~\ref{magnetization} shows the order parameter $m$ as function of
$(\rho_1-\rho_1^c(\rho_3))/\rho_1$ for $\rho_3=0.2$, 0.1, and
0.05. Only those data points are shown that are not affected by
finite-size effects. (When finite-size effects are present, the
fraction of flipped spins decreases with increasing system size. The
reason is that only the area behind the interface was evaluated (not
taking into account the first $L/2$ columns), which cannot contain
unflipped regions larger than $L$ and has therefore less unflipped
spins for smaller system size.) With the maximum system height $L=3200$
used in the simulations, the depinning threshold could not be 
approached closer than shown in figure~\ref{magnetization}. 
As the log-log plot shows, the data points can be fitted within the 
error bars by a power law
\begin{equation}
\label{m-scaling-rho1}
m \propto ((\rho_1-\rho_1^c(\rho_3))/\rho_1)^\beta
\end{equation}
with an
exponent $0.076 \le \beta \le 0.124 $ that takes decreasing values for
decreasing $\rho_3$ and is smaller than the percolation value
$\beta_{\text{perc}}=5/36 \simeq 0.139$. Note that these results have
to be treated with caution: (i) The data points cover less than a
decade. (ii) Although the fitted power laws lie within the error bars,
there appears to be a slight increase in slope with decreasing
$(\rho_1-\rho_1^c(\rho_3))/\rho_1$ for all three data sets. (iii) The
data are taken relatively far away from the critical point $m=0$, in
fact probably already outside of the scaling regime.  (Attempts to
obtain scaling collapses of the data using the general scaling form
given above did not work very well.) The exponent $\beta$, is actually
only defined near $m=0$ as as $$[\text{d} \ln m /\text{d}
\ln((\rho_1-\rho_1^c(\rho_3))/\rho_1)]_{m=0}\, .$$ (iv) The simulated
system is rather small. It is known from other nonequilibrium systems
with quenched disorder\cite{per97} that finite size effects tend to be
rather large and result in somewhat shifted values for the critical
exponents. Curiously, in \cite{per97} collapses for the magnetization
curves seemed also to be the hardest to obtain.  (v) With increasing
$(\rho_1-\rho_{1}^{c})$, the exponent $\beta$ may cross over to some
other value $ \beta'$. In fact, our scaling theory (see section
\ref{scaling-theory}) suggests that in the low disorder regime, very
close to depinning, ( $(\rho_1-\rho_{1}^{c}(\rho_3))/\rho_1 \ll
(\rho_1^c)^{x/z}$, where $x/z$ is a universal exponent estimated
in section \ref{scaling-theory}),
one has $m \sim
((\rho_1-\rho_{1}^{c}(\rho_3))/\rho_1)^{\beta_{\text{perc}}}$ with
${\beta_{\text{perc}}} = 5/36$; and further away ( $(\rho_1^c)^{x/z} \ll
(\rho_1-\rho_{1}^{c}(\rho_3))/\rho_1$ ), one has $m \sim
((\rho_1-\rho_{1}^{c}(\rho_3))/\rho_1)^{\beta'}$ with $\beta'=0$. Such
a scenario would explain the apparent decrease of $\beta$ with
decreasing disorder.
\begin{figure} \narrowtext
\vskip 1cm 
\centerline{\psfig{file=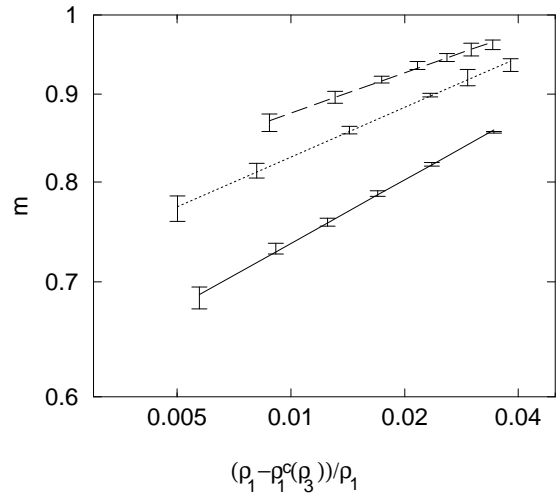,height=2.5in,angle=-90}}
\caption{The fraction of spins flipped by the avalanche as function of
the distance from the critical density $\rho_1^c(\rho_3)$, for
$\rho_3=0.2$ (solid), 0.1 (dotted), and 0.05 (dashed) for a system height
up to $L=3200$. The measured
data points sit in the center of the error bars, and the error bars
have a length of two times the standard deviation. The lines are a
power-law fits to the data, with the exponents $\beta=$ 0.124, 0.096, and
0.076. (Note that for percolation ${\beta_{\text{perc}}} = 0.139$.) }
\label{magnetization} 
\end{figure}
Below, in section
\ref{analytical}, we will give analytical arguments that for small enough 
$\rho_1-\rho_1^c$ the exponent $\beta$ is identical to
the percolation exponent. Furthermore, using a test
simulation of another system for which $\beta$ is known to be equal to 
$\beta_{\text{perc}}$, we will show
that corrections to scaling tend to bias the numerically
observed value for $\beta$ towards a ``wrong'' value 
$\beta < \beta_{\text{perc}}$.

Nevertheless, a few valid conclusions can be derived from the
numerical data. First, at least for $m > 0.5$ there seems to be no
tendency to approach a finite saturation value of $m$. If this
tendency continues for smaller values of $m$, it indicates a
continuous depinning transition.  Second, the
critical interval $\rho_1-\rho_1^c(\rho_3)$, for 
which $m$ is smaller than some
threshold value (e.g., 0.5), becomes smaller with decreasing $\rho_3$. Our
conclusion will be that for sufficiently small  $\rho_1-\rho_1^c$
the exponent $\beta$ does not decrease with
$\rho_3$, but that the {\it amplitude} of the power law has to increase with
decreasing $\rho_3$.  Below, the size of the critical region will be
estimated using two different arguments, leading to a power-law
divergence of the amplitude as $\rho_3 \rightarrow 0$. 
Third, using finite-size
scaling, the fractal dimension of the invaded region can be estimated.
The result is compatible  with the percolation value
$D_f=91/48 \simeq 1.896$, as found earlier in \cite{ji91,koi92}. 

\subsection{The unflipped regions left behind} \label{loecher}
Figure~\ref{holes} shows the size distribution $n(s)$ of unflipped domains
behind the interface for $\rho_3=0.2$ and different values of $\rho_1$
close to $\rho_1^c$. The
size is defined as the number $s$ of unflipped spins within the
domain. Unflipped spins that are nearest or next-nearest neighbours
are defined to belong to the same domain. (This definition allows for 
system spanning clusters of unflipped spins coexisting with 
system spanning clusters of flipped spins at the depinning point).
\begin{figure} \narrowtext
\centerline{\psfig{file=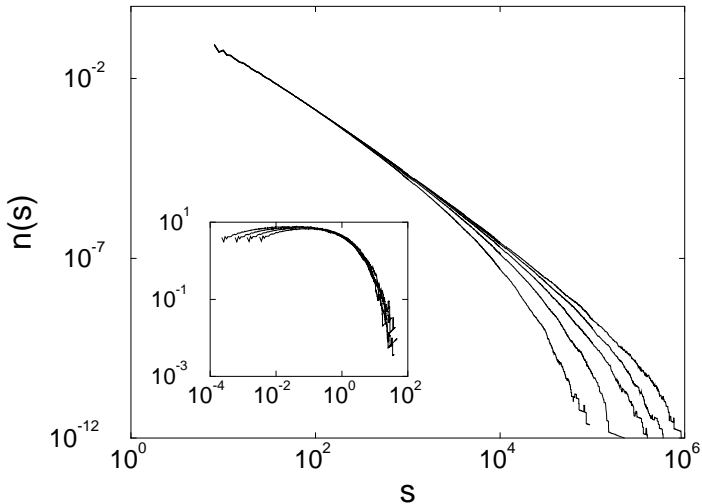,height=3.0in,angle=-90}}
\vskip 1cm
\caption{Size distribution of unflipped regions for $\rho_3=0.2$ and 
$\rho_1=0.4385$, 0.44, 0.442, 0.445, and 0.45 (from right to left). 
$\rho_1^c(\rho_3=0.2) = 0.4345 \pm 0.005$. The inset shows a collapse
of these data, plotting $s^\tau n(s)$ versus 
$((\rho_1- \rho_1^c)/\rho_1)^{\sigma_n} s $ with $\tau = 1.85$, 
$\sigma_n = 1.8$, and $\rho_1^c = 0.437 $.
\label{holes}}
\end{figure}
Near depinning we tested the scaling ansatz $n(s) \sim 1/s^\tau
f_n(((\rho_1- \rho_1^c)/\rho_1)^{\sigma_n} s) $ with $f_n$ an
appropriate scaling function. This ansatz seems to give a collapse for
$\tau=1.85$, $\sigma_n=1.8$, and $\rho_1^c=0.437$ (see inset). This
apparent scaling behaviour, however, cannot hold for very small
$(\rho_1-\rho_1^c)/\rho_1$, since the integral $\int_1^\infty sn(s)
ds$ must be normalized to 1, allowing either for a scaling form with
$\tau=2$, or for $\tau<2$ with no universal scaling behaviour.
We can
rule out the possibility $\tau>2$, since the simulations as well as
the analytical arguments below suggest that the total area of small
unflipped regions decreases at the expense of large unflipped regions
as $\rho_1$ approaches $\rho_1^c$. A value $\tau>2$, in contrast,
would imply that the area fraction covered by large unflipped regions
is negligible.

Indeed, with decreasing distance from the critical value
$\rho_1^c(\rho_3)$, the curve becomes flatter and does not converge to
an asymptotic curve with finite cutoff.  This behaviour can be
interpreted as another indication that the order parameter vanishes
when the critical point is approached.  If it did not vanish, the size
distribution of unflipped domains would approach some limit
distribution with a finite cutoff at the critical point. Why this does
not happen for a vanishing order parameter, is best illustrated for
the site percolation limit of large disorder, $\rho_1=p,\,
\rho_4=1-p$. There, the invaded area is identical (except for the
first few columns) to the infinite percolation cluster. Clearly, as
$p$ is decreased towards its critical value $p_c \simeq 0.59$, larger
and larger branches of the infinite percolation cluster become
disconnected from it and are therefore no more flipped. All the
unflipped domains formerly contained within this branch fuse, and
larger unflipped domains are formed at the expense of smaller domains.

\section{Analytical results} \label{analytical}

\subsection{The exponent $\beta$}
The simulation results shown in figure~\ref{magnetization} give a 
value of $\beta$, somewhere between $0.076$ and $0.124$, depending
on $\rho_3$. The obtained range does not include the percolation
value, which is ${\beta_{\text{perc}}} = 0.139$. As argued in
subsection \ref{OP}, the data however are not conclusive, and a
universal value $\beta=\beta_{\text{perc}}$ (or a close value) 
cannot be ruled out.  A value $\beta$ different from 
$\beta_{\text{perc}}$, would mean that
any deviation of the densities $\rho_n$ from the
percolation values should be a relevant perturbation of percolation
theory. In other words, the conditional flipping of spins depending on
the number of flipped neighbours (i.e.~ $\rho_2 \neq 0$, $\rho_3 \neq
0$), should change the universality class. 
The case that $\beta$
would depend on $\rho_3$ is highly unlikely. It would imply
that even the extent to which spins are flipped
as function of the state of their neighbours, would affect the value of
the critical exponent. 

In the following we give several points on the critical surface apart
from $\rho_1=0.59..$, $\rho_4=1-\rho_1$ where we can show by an
analytic mapping onto percolation models that
$\beta=\beta_{\text{perc}}$. These results will provide a strong case
that $\beta$ is indeed universal.  One such point is given by
$\rho_1=p_b,\, \rho_2=p_b(1-p_b),\, \rho_3=p_b(1-p_b)^2,
\rho_4=(1-p_b)^3$, with $p_b=p_b^c=1/2$ being the critical threshold
for bond percolation on a square lattice. In order to understand this,
consider a bond percolation problem, where a pair of neighbouring
sites is connected by a bond with probability $p_b$.  We start with a
row of up-spins at one end of the system and allow the avalanche to
proceed to any site that can be accessed via open ({\it i.e.}
existing) bonds.  Some sites are invaded at the first
nearest-neighbour contact with the spin-flip avalanche, others at
the second contact, still others at the third contact, and the rest
not even at the third contact. Obviously, the proceeding avalanche
cannot distinguish whether it moves through a bond percolation system
or a system with sites of different ``colours'' that are assigned
according to the probabilities $\rho_n$ (Eq.~(\ref{rho})). Clearly, the
invaded bonds will form a bond percolation cluster. 
Since bond percolation can be
mapped onto site percolation on a different lattice\cite{sta94}, and
since the site percolation critical exponents  
do not depend on the lattice type, the number of invaded bonds
diverges as $n_b \approx C (p_b-p_b^c)^{\beta_{\text{perc}}}$. In
\cite{ble76}, it is proven that the number of invaded sites in a bond
percolation problem increases with the same exponent as the number of
invaded bonds. Consequently, the order parameter exponent is identical 
to $\beta_{\text{perc}}$ for the above choice of the $\rho_n$. 

\begin{figure} \narrowtext
\centerline{\psfig{file=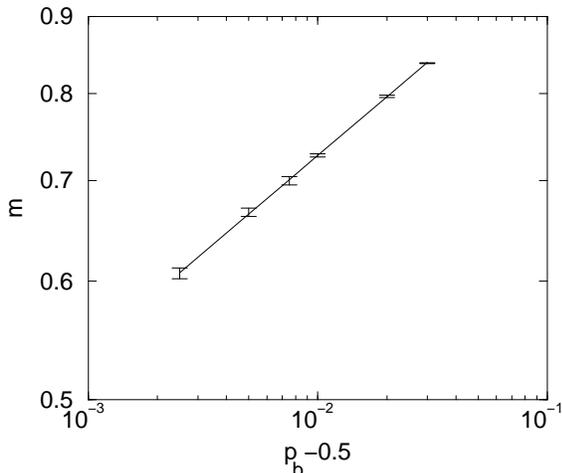,height=2.5in,angle=-90}}
\caption{Fraction of invaded sites for bond percolation with $L=3200$,
from a simulation with $\rho_1=p_b,\, \rho_2=p_b(1-p_b),
\rho_3=p_b(1-p_b)^2,\, \rho_4=(1-p_b)^3$, with $p_b=p_b^c=1/2$.
The solid line is a power law fit with the exponent $\beta =0.1296$.}
\label{bondperc}
\end{figure}
Figure~\ref{bondperc} shows the number of invaded sites as function of
$p_b-p_b^c$. The simulations were performed 
by setting $\rho_1=p_b, \,\rho_2=p_b(1-p_b), 
\rho_3=p_b(1-p_b)^2,\, \rho_4=(1-p_b)^3$, with $p_b=p_b^c=1/2$ in the 
previous simulations. The solid line is a power-law fit to the data, 
with an exponent $\beta=0.1296$, which is smaller than the true exponent
$\beta_{\text{perc}}=5/36\simeq 0.139$ which we obtained from
analytical arguments. This suggests that corrections to scaling
modify the asymptotic power law to an apparently different power law
further away from the critical point. Not shown in the figures is the
result for the fractal dimension of the interface, which is again
$4/3$, as in the previous simulations. We therefore have strong
reasons to believe that the bond percolation case discussed in this
subsection belongs to the same universality class as the simulations
described in the previous section.

For other special values of the densities $\rho_n$, mappings on
various other types of percolation are possible. For example,
site-bond percolation (i.e., bond percolation where only the fraction
$p_s$ of all sites are accessible) interpolates between site
percolation and bond percolation and is obtained for the densities
$\rho_1=p_s p_b $, $\rho_2=p_s p_b (1-p_b)$, $\rho_3=p_s p_b (1-p_b)^2$,
and $\rho_4=p_s (1-p_b)^3+(1-p_s)$. Mappings on short-range correlated bond 
percolation are also possible, where the probability that a bond
is open depends on the number of bonds pointing to the same site
(in this situation, bonds must be given an orientation). 

To summarize, there is strong evidence that the critical exponent
$\beta$ is indeed universal and identical to its
percolation value $5/36$. 

\subsection{The size of the critical region}
As shown in figure~\ref{magnetization} and mentioned in section \ref{OP},
the parameter interval during which the order parameter increases from 
zero to a given finite value becomes smaller with decreasing
disorder strength. Together with the results of the previous section,
this suggests the form 
(for $(\rho_1 - \rho_1^c)/\rho_1 \ll (\rho_1^c)^{x/z}$ )
\begin{equation} \label{kappa}
m \simeq C \rho_3^{-\kappa} ((\rho_1-\rho_1^c(\rho_3))/\rho_1)^\beta\, ,
\end{equation}
with $\beta = {\beta_{\text{perc}}}= 5/36 \simeq 0.139$, and with some exponent
$\kappa$.  In the following, we will first derive an estimate of the
size of the critical region and of $\kappa$ based on the understanding
that the invaded area is a coarsened percolation cluster, and then a
different estimate that is based on a study of the unflipped
domains. Both estimates agree within 12 percent, suggesting that the
true value is of the same order as the estimated values. 

\subsubsection{Arguments from percolation theory}

From the previous section, we know that the invaded area can be viewed
as an infinite percolation cluster with additional sites added to it.
Since a spin-flip avalanche can only be stopped by ``grey'' sites
that occur with a probability $\rho_3^{-1}$, we assume now that to
each site of the infinite percolation cluster all sites within a
distance $\rho_3^{-1}$ are also flipped. This assumption is supported
by the result in \cite{koi92} that the ``finger width'' diverges
roughly as $\rho_3^{-1}$.
Clearly, when this distance
becomes larger than the percolation correlation length $\xi$, practically
all spins are flipped, and the system is no more in the critical
region. We therefore find
$$((\rho_1-\rho_1^c(\rho_3))/\rho_1)^{-\nu}\propto \rho_3^{-1}$$
at the boundary of the critical region, with $\nu = 4/3$ known from
percolation theory, leading to a size of the critical region
\begin{equation}
((\rho_1-\rho_1^c(\rho_3))/\rho_1)\propto \rho_3^{3/4}\, . \label{estimate1}
\end{equation}
At the boundary of the critical region, $m$ has some finite value, and setting 
$m$ constant in Eq.~(\ref{kappa}), we find $\kappa/\beta=3/4$, or
\begin{equation}
\kappa=3\beta/4=5/48 \simeq 0.104 \label{kappa1}\,.
\end{equation}

\subsubsection{Arguments using unflipped domains}

Let us now consider the unflipped domains left behind by the infinite
avalanche, and let us measure all distances in units of the ``step
size'' $l \simeq 1/\rho_1$. The probability that a (pinning) path of a 
given step number $k$ has its final point within unit distance $l$ from 
the initial point is independent of the step size $l$ for large $l$. Since
there are of the order $l^2$ sites within distance $l$ from the
initial point, the probability that a path of $k$ steps forms a closed
loop vanishes as $l^{-2}$, which is proportional to $ \rho_3$. On the
other hand, there is a cutoff $1/(1- p_{surv}/\rho_1)$
to the number of steps of a pinning path (see  Eq.~(\ref{survival})),
leading to
$$k_{\text{max}}\propto [(\rho_1-\rho_1^c(\rho_3))/\rho_1]^{-1}\, $$ 
for small $\rho_3$. 
As long as there
exists only small and rare unflipped domains, the system is beyond the
critical region. The critical region can be characterized by the
condition that the distance covered by a pinning path,
$k_{\text{max}}^{3/4}$, becomes of the same order as the distance
between unflipped domains. Then, the picture of rare independent
unflipped domains is no longer valid, since these domains can be
connected by pinning paths, leading to a divergence of the size of
unflipped domains. The above condition reads 
$$k_{\text{max}}^{3/4} \propto l \propto \rho_3^{-1/2} \,  ,$$ 
and gives the following estimate of $\kappa$  by again 
putting $m=\text{const}$ and using the assumption of Eq.~(\ref{kappa})
\begin{equation}
\kappa=2\beta/3 = 5/54 \simeq 0.093 \,. \label{kappa2}
\end{equation}

This argument also shows that the size distribution of unflipped
domains does not converge towards a fixed function with finite cutoff
for $\rho_1\to
\rho_1^c(\rho_3)$, but rather that small unflipped domains become
connected to form large unflipped domains, as observed in the
simulations. 

\subsection{Scaling theory}
\label{scaling-theory}
The results for the order parameter and the unflipped domains
lead  to a scaling theory that is
similar in spirit to the one proposed in \cite{bra85} for the
equilibrium random-field Ising model in two dimensions
at low disorder. We introduce
the scaling variable
$$h = (\rho_1-\rho_1^c)/\rho_1, \, $$ 
which measures the distance from
the depinning threshold and is roughly equivalent to $(H/R-H_c/R)$, and
$$t=\rho_1^c\, ,$$ 
which is a measure for the disorder strength and is
to leading order 
$$\rho_1^c \simeq \sqrt{\rho_3} \propto \exp[-0.343J^2/R^2]\,.$$ 
The
characteristic length scale $t^{-1}$ is the length of straight
interface segments. The correlation length $\xi$ is most naturally
identified with the diameter of the largest unflipped domain, weighted by the 
density of these domains, and can
be expected to scale with some power of $t$. In equilibrium, the
correlation length scales also as $\exp[CJ^2/R^2]$, with a constant
$C$ different from the depinning problem.

Under coarse graining, the scaling variables change to 
$$ h'=b^xh \qquad t'=b^zt.$$ 
We assume that the exponent $x$ is $x=d-d/2=1$, because
the renormalized external field $H$ and random fields $\{h_i\}$ are
given by the sum of the corresponding fields in the cell, leading to
dimensions $d$ and $d/2$ respectively \cite{bra85}. 
Here, the main assumption is that we can apply equilibrium
rescaling under coarse graining to this essentially non-equilibrium
problem. The reasoning is that in the limit of low disorder almost
all (except for a vanishing fraction as $R \rightarrow 0$)
coarse grained ``block spins'' flip {\it coherently} thereby optimally
lowering their local energy, and obeying essentially the same rules
as single spin flips on shorter length scales.
The exponent $z$ will be related below to the exponent $\kappa$. The
correlation length $\xi$ and the order parameter $m$ transform under
coarse graining according to
$$\xi'=\xi(t',h')=b^{-1}\xi(t,h)$$
and 
$$m'=m(t',h')=b^{y}m(t,h)$$ with $y=0$ (since all spins within a box
of size $b^2$ are parallel at $t=0$), leading to
$$\xi \simeq t^{-1/z}\tilde \xi(h/t^{x/z})$$ and
 $$ m \simeq t^{y/z} \tilde m(h/t^{x/z}).$$
As is seen from figure \ref{flowdiagram}, on long length scales 
the system flows to the percolation fixed point at infinite disorder.
The zero disorder fixed point and the percolation fixed point
are connected by the depinning
critical line which is described by the correlation critical
fixed points on sufficiently long length scales. 
In the following we use information about the percolation fixed point 
to extract the asymptotic behaviour of the scaling functions as well.
For $h \ll t^{x/z}$, we expect  
$$ \tilde \xi \sim (h/t^{x/z})^{-\nu}$$
with $\nu=\nu_{\text{perc}}=4/3$, or 
$$ \xi \propto h^{-\nu}t^{(x\nu-1)/z}\,.$$ For smaller disorder, $\xi$
is also smaller (for the same value of $h$), in agreement with our
previous finding that the width of the critical region becomes
smaller. 

From the previous section we know that for $h \ll t^{x/z}$
$$\tilde m (h/t^{x/z}) \sim (h/t^{x/z})^{\beta}$$
with $\beta = \beta_{\text{perc}}$. 
This gives for $h \ll t^{x/z}$
$$m \sim t^{(y-\beta x)/z} h^\beta$$
and the scaling relation
$$ \kappa=(\beta x -y)/2z.$$ Inserting the known values for the
exponents $y$, $x$, and $\beta$, the relation between $\kappa$ and $z$
becomes
$$z=5/72\kappa\, .$$
The two above estimates for $\kappa$ give then $z=4/7$ or $z=9/14$.

In the opposite limit $h \gg t^{x/z}$, the order parameter saturates
at 1, i.e., $$\tilde m (h/t^{x/z}) \sim (h/t^{x/z})^{\beta'}$$ with
$\beta'=0$. For intermediate values of $h/ t^{x/z}$, the scaling
function $\tilde m$ interpolates between the two limits. The exponent
$\beta'$ is also observed for faceted growth, where never more than
two ``colours'' are present, and where depinning occurs at $\rho_2=1$.
In the flow diagram, faceted depinning occurs at the left-hand fixed
point, and the flow follows the perpendicular axis. 

The correlation length $\xi$ is close to zero for $h \gg t^{x/z}$, since
essentially no unflipped domains are left behind. This means that the
scaling function $\tilde \xi$ becomes proportional to
$(h/t^{x/z})^{-\nu'}$, with $\nu'=\infty$.

\begin{figure} \narrowtext
\centerline{\psfig{file=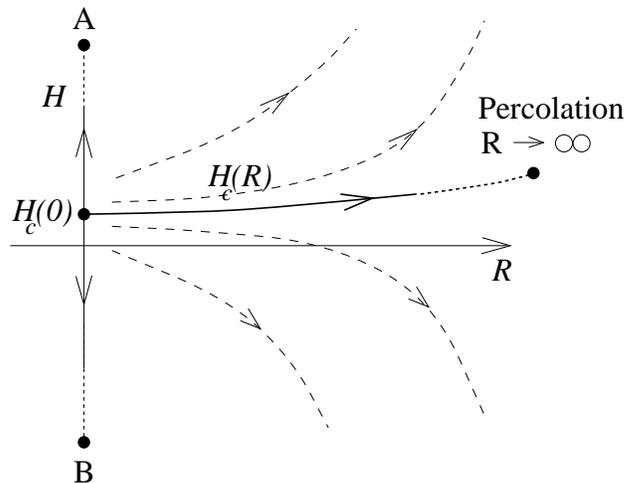,height=2.5in,angle=-90}}
\caption{Sketch of the flowdiagram for the order parameter.
There are 4 fixed points in this diagram: 
at $R=0$, $H=H_c(0)$ (with $H_c(0) = 1.172 J$
for the model discussed here), which is the zero disorder fixed point 
discussed in this paper; at finite $H$, $R \rightarrow \infty$, which is 
the percolation fixed point; and at $H = \pm \infty$, $R = 0$ (A and B),
which attract all the flow above respectively below the critical line.
These to fixed points correspond to a completely flipped system and to a
system that is not invaded at all.
The thick line marked $H_c(R)$ is the depinning line.
The arrows indicate the direction of flow under coarse graining.
The diagram explains why systems very close to $H_c(R)$ are dominated
by percolation critical exponents on long length scales, as discussed in
the paper.}
\label{flowdiagram}
\end{figure}

\section{Conclusions} \label{conclusions}
 
The results of this paper confirm that 2 is the lower critical
dimension for the transition from percolation-like to conventional
depinning of a domain wall in the random-field Ising model. This
conclusion is based on the numercial observation that the order
parameter (the fraction of flipped spins) vanishes continuously as the
depinning threshold is approached, even in the limit of very small
disorder. Numerical results as well as analytical arguments show that
the size distribution of unflipped domains left behind by the
spin-flip avalanche becomes flatter closer to the depinning
threshold, not allowing for a nonvanishing order parameter at the
threshold. 

Furthermore, this paper supports the hypothesis that the order
parameter exponent $\beta$ is the same as in uncorrelated site
percolation. Since numerical data are not conclusive and rather
indicate an exponent $\beta$ that depends on the disorder strength, an
explicit mapping of the infinite avalanche of the depinning problem
onto an infinite cluster in percolation theory is performed for
several distinct parameter values. The reason why the asymptotic value
of the critical exponent $\beta$ cannot be seen in the simulations is
that the width of the critical region shrinks to zero as the disorder
strength vanishes. Analytical arguments in this paper estimate the
value of the exponent that characterizes the width of the critical
region, and lead to a scaling theory that relates this exponent to
other exponents.

As argued in \cite{nar93}, depinning in the random-field and
random-bond Ising models belong to the same universality class,
and the non-equilibrium
random bond and random field Ising model have the same symmetries
near the critical point (for the same reasons as given in \cite{dah96}),
we therefore expect that the results of this paper are also valid
for the random-bond Ising model.
In contrast, the equilibrium
critical behaviour of the two models is different.

The question of the universality of the exponent $\beta$ occurs also
in the context of bootstrap and diffusion percolation \cite{adl90},
where all unoccupied sites of a site percolation problem that have a
certain number of occupied neighbours are also occupied. Recently,
evidence was found that the exponent $\beta$ is universal in two
dimensions \cite{cha95}.

The scaling theory presented in this paper discusses only the order
parameter and correlation length in the bulk, after the spin-flip
avalanche has transversed the system. The scaling behaviour of the
front is somewhat simpler for small disorder, since for small $t$ and
$h$ the correlation length is identical to the length of a pinning
path, which is proportional to
$$(l k_{\text{max}})^{3/4} = (th)^{-3/4},$$ leading to a scaling variable
$th$. A pinning path is essentially a self-avoiding random walk.
Therefore the fractal dimension of the front is 4/3, which is
different from the percolation value 7/4. However, close to the site
percolation fixed point (i.e. for large disorder), a crossover between
the two exponents should be observed. Curiously, this implies that the
flow diagram for the correlation length of the front shows a flow from
the percolation fixed point to the $t=h=0$ fixed point, which is the
opposite direction to the flow in figure~\ref{flowdiagram}. A scaling
theory for the front should therefore be performed in the
neighbourhood of the percolation fixed point, which was not the focus
of this paper.  Other front properties like the size distribution of
avalanches below the depinning threshold and the velocity of the front
were not studied in this paper either and have still to be determined.

It is certainly possible to generalize the scaling theory of this
paper for the transition from percolation-like to conventional
depinning to the neighbourhood of two dimensions by performing a
$2+\epsilon$ expansion, in a way similar as in \cite{bra85} for the
equilibrium model. An expansion around the upper critical dimension is
a bigger challenge. Since in dimensions larger than 2 the disorder
strength is not vanishingly small at the phase transition between the
two different modes of depinning, the neglection of spontaneous spin
flips away from the domain wall is realistic only under certain
circumstances, for example for
fluid invasion, for magnetic samples in a gradient field, or 
in the presence of certain long range interactions\cite{dah96}.
If, on the other hand, one includes these spontaneous spin flips, one
arrives at a hysteresis model for which an expansion arond the upper
critical dimension 6 was successfully performed in \cite{dah96}.

\acknowledgements 

We would like to thank A. Bray, J. Essam, T. Prellberg,
and in particular J.P. Sethna for helpful discussions. 
This work was supported by EPSRC Grant No. GR/K79307,
the Society of Fellows of Harvard University,
and NSF via DMR 9106237, 9630064, and Harvard's MRSEC.

\narrowtext

\end{multicols}


\begin{thebibliography}{5}
\bibitem{footnote2} $R_c = 0$ needs 
not necessarily be true for a lower critical dimension.
The Kosterlitz Thouless transition\cite{Kosterlitz}, for example
occurs at the lower critical dimension at a nonzero critical 
temperature.
\end{thebibliography}
\end{document}